# Ultrasound vibrations of plant cells membranes: water lift in trees, electrical phenomena

Mark E. Perel'man[*)] and Galina M. Rubinstein
*(Racah Institute of Physics, Hebrew University of Jerusalem, 91904, Israel)*

ABSTRACT
Alterations of charges density on membranes (potentials) of plant cells by ion currents lead to disbalance of Coulomb forces compressing membranes and counteracting Hookean elastic forces. It should results in membrane vibrations till establishment new equilibrium with generation of ultrasound and electromagnetic radiations. Such vibrations lead to acoustical flows, power of which is sufficient for water lift in trees, its degassing, for cycloses and some electric phenomena in plants and so on. This phenomenon explains the known observations of ultrasound noises in trees and redistribution of particles in the phloem flow. Known supervisions and consideration of anatomy of plants result in estimations of frequencies of the order of 150-200 kHz for acoustical flows in xilem and three times bigger in phloem. The checking of offered mechanism can be carrying out via investigation of acoustic and electromagnetic radiation of plants.

## 1. Introduction and qualitative description

Some basic phenomena of plant physiology, that must be described by physics or, at least, coordinated with it, have not up till now the common explanation. Among such phenomena are hydraulics of trees (xylem transport and phloem flow), cyclosis, i.e. the movements of cytoplasm within a plant (or even animal) cell, deletion of emboling gas bubbles and bypass of defeat sections at water transport, the origin of electrical potentials and their distribution and variation over tree, e.g. [1].

So, the lifting of water in trees up to height 100 m demands search of especial pumping effects. The explanation of this lifting via cohesive features of pure water was offered about century ago by Dixon and Joly [2] and was criticized by many experts. We do not reiterate their arguments, which are comprehensively set out e.g. in the review article [3] with extensive literature references (note only cavitation phenomena in vessels, that completely discredit the cohesive hypothesis, e.g. [4, 5]).

We shall consider such model which can explain some phenomena of plants physiology: alteration of potential of charged cell membranes induce their mechanical vibrations together with cell walls, i.e. generate acoustic waves, whose frequencies are in the ultrasound region and radiated power of a lot of cells become sufficient for observed effects. (The hypothesis of ultrasound role in water transport had been announced in [6, 7], but without revealing of their origin.)

Physical foundation of this effect is such. Both sides of some part of membrane with more or less chaotically distributed charges represent the electrical capacitor or the double electric layer. This transmembrane potential can be different in different parts of cell and altered with time (ion currents and so on). In such "structure" Coulomb forces pulls plates together and compresses its content, but the elastic Hookean forces of membrane withstand them and together they maintains an equilibrium. If charge densities, i.e. the transmembrane potential will be changed, it will upset the balance of powers and this part of membrane will begin to vibrate with damping till establishment of new equilibrium state: these processes can be considered as the parametric oscillations, which generate acoustic (in our case ultrasound) and electromagnetic waves of the same frequencies.

---

[*]). Correspondence and requests for materials should be addressed to E-mail: mark_perelman@mail.ru

The ultrasound radiation (USR) from different cells or even from parts of single cell would be independent, i.e. these vibrations will be integrally resulted as an acoustic noise with some possible distribution of frequencies and so on, but the noise of sufficient power is enough for considered phenomena. Such features complicate its direct registration, but as the considered processes should be accompanied by electromagnetic radiation on the same frequencies, the registration should be possible by more apprehensible electromagnetic devices.

At dissipation of acoustic wave in medium its energy is thermalized, but its momentum must lead to motion of some constituents of medium and in a liquid substance it generates so named acoustic flow [8]. Acoustic flows have the unique peculiarity: more dense constituents acquire bigger momentum and can outstrip liquid, in which they move. Just such type of motion was observed in [9]: in the horizontal vine big particles (radoactively labelled glizins, xylose particles, vitamins) moved faster than phloem flux which carrier them.

The USR oscillations appearing in the root cells will propagate into xylem vessels where the resistance of water flow is considerable lower than at any other side: capillar columns existing in them play a role of directed acoustoguides.

Let's briefly mention, before numerical estimations, some qualitative features of proposed USR flows. USR waves must lead to degassing of xylem flow; remember that the problem of embolie is a critical one for the Dixon- Joly approach. Near to a place of damage of xylem channel this acoustic flow will redirect water to bypass this place by leakage through cell walls into other channels. At trees felling their root system some period continue work, i.e. generate USR, but wave reflected from the exterior surface will interfere with initial wave that leads to their attenuation; therefore the "weeping" of trees is comparatively weak. Among other properties it is known that USR has thixatropic activity: it leads to comparative liquefaction of gum into living trees and to possibility of its deallocation.

We shall begin with consideration of vibrations of membranes, together with their walls, at alteration of charges density that result in corresponding acoustic and electromagnetic radiations (Section 2, another approach is given in the Appendix). In the Section 3 some simple heuristic estimation of frequencies limits based on direct usage of observations data are considered; whereupon the more strict calculations with taking into account anatomy of water and sap conducting vessels are performed. It leads also to some estimation of flows velocity and power.

In the Section 4 the electrical phenomena accompanying acoustic flow are considered, it gives, in particular, additional possibility for estimation of the main magnitude, the frequency of USR. In the Section 5 are enumerated some different questions related to the offered theory. The main results are briefly summed in the Conclusions.

## 2. Membrane vibrations

For estimations of forces and frequencies such parameters of cells will be used [10]:
Standard thickness of biomembrane $h \sim 10$ nm;
Thickness of plant cell wall $H \sim 1 \div 10$ mcm;
Mean electrical potential on membrane is of order $U \leq 0.1$ V;
Specific surface capacity $C/S = 1$ pF/mm$^2$;
Electrical resistance of breakdown $R/S = 10^2 \div 10^5$ Ω/cm$^2$;
Mechanical properties are determined by the Young's modulus $Y$: the longitudinal modulus $Y' \sim 10^9$ Pa, but for transversal forces it must be sufficiently lower and $Y \sim 10^6 \div 10^7$ Pa can be taken.

Condition of equilibrium, i.e. an equality of the Coulombian and Hookean pressures,

$$p_{elast} = p_{electr} \qquad (2.1)$$

at membrane pressing on the depth $\Delta h$ must be written as

$$\frac{C}{S}\frac{U^2}{h} = Y\frac{\Delta h}{h}. \qquad (2.2)$$

The magnitude of this pressure for taken data is of order $p \approx 1$ Pa.

At a charge breakdown on the section $S$ the membrane and its connected wall come into motion together, i.e. the oscillating mass $m = \rho S H$, where $\rho$ is their mean density. Therefore the equation of motion at action of electric force becomes:

$$m\frac{d^2\Delta h}{dt^2} = Y\frac{\Delta h}{H}. \qquad (2.3)$$

Its solution in the form $\Delta h = A \sin(2\pi f\, t)$ directly leads to an estimation of frequency:

$$f = \frac{1}{2\pi H}\sqrt{\frac{Y}{\rho}} \sim 0.5 \div 1.5 \text{ MHz}, \qquad (2.4)$$

this is close to values estimated below by other methods.

Let's estimate parameters of electrical current at such breakdown. Every elementary charge (ion channel) is located on an area of order $S_1 \sim 3 \div 5$ nm$^2$. Thus at simultaneous breakdown of section $S \sim 100 \div 1000$ nm$^2$ through membrane can be transferred current $I \approx ef(S/S_1) \sim 1 \div 10$ pA, here an estimation (2.4) was used.

The frequency of oscillations generated by charges transition can be independently estimated via velocity of ions transverse diffusion through membrane. The coefficient of diffusion is of order $D \approx h^2 f/2$. As at the ion transport $D = (1 \div 5)\, 10^{-7}$ cm$^2$/s [11], it leads to
$$f' \sim 2D/h^2 \sim 0.2 \div 1 \text{ MHz}, \qquad (2.5)$$
very close to (2.4).

Now we must estimate the energy of generated waves and its possibility to powered, for example, cytoplasm at cyclosis.

The energy of ultrasound vibration
$$W_{USR} = p^2/\rho c^2, \qquad (2.6)$$
where $c = 1.5 \cdot 10^5$ cm/s is the sound speed in water. As the velocity of cytoplasm streaming $v \sim 10$ mcm/s (e.g. [11]), the equaling of (2.6) with kinetic energy of powered area of liquid
$$W/S = mv^2/2S = \rho L v^2/2,$$
where $L$ is the depth of powered mass, leads to $L \sim 0.1$ mm, which seems non contradictory and can explain some manifestations of cyclosis.

## 3. Water lift and phloem flow

There are several possibilities for estimation of US frequencies on base of plant physiology. We shall begin with the simplest ones.

### 3-1. ESTIMATIONS BY OBSERVATIONS DATA

At flow of viscous liquid through channel the inhibition layer of depth $\delta \sim (\eta/\omega\rho)^{1/2}$ near walls exists, where $\eta, \omega, \rho$ are dynamical viscousity, circular frequency and liquid density, correspondingly. As the depth of this layer must be essentially lesser the radius $R_0$ of conducting channels, we receive the condition:

$$\nu \gg \nu_{min} \approx \eta/2\pi\rho R_0^2 \sim 3\cdot 10^4 \text{ Hz}. \qquad (3.1)$$

We used for this estimation such data: $R_0 \sim 10^{-3}$ cm, $\eta \sim 10^{-2}$ Ps for water and $\eta \sim 10^{-1} \div 1$ Ps for cytoplasm in cell.

If the conducting channel is represented as an ideal circular cylinder of diameter $d \sim (3 \div 8) \cdot 10^{-3}$ cm with rigid walls, then, as it is well known from the theory of waveguides, plain waves can propagate in the axial direction at frequencies:
$$\nu < \nu_{10} = 1.84c/2\pi d \sim 20 \text{ MHz}. \tag{3.2}$$

More one simple estimation is connected with possible degassing of lifted flux. The degassing is maximal, as is known from the USR practice, at
$$\nu_{\deg as} \sim 200 \text{ kHz}. \tag{3.3}$$

From the heuristic principle of biological expedientness it can be expected that really induced frequencies would be close to (3.3).

## 3-2. ELECTROACOUSTIC ANALOGIES

All these estimations can give some very preliminary heuristic values and therefore more strict examinations are needed.

In many articles on plant physiology the analogy with the Ohm law at consideration of liquid flows is used. But real situation is much complicate and requires the consideration of electroacoustic analogy with electric circuits of alternating current (e.g. [12]).

Let us consider such analogies. With tracheid cell can be compared the acoustical capacity
$$C_A = SL/\rho_{liq}c^2, \tag{3.4}$$
where $S$ and $L$ are the cross-section and length of cell, $\rho_{liq}$ is the density of liquid. With perforated hole of radius $a$ and length $l$ is compared the acoustical inductivity
$$L_A = \frac{\rho_{liq}}{\pi a^2}(l + 0.8a\sqrt{\pi})^{1/2}. \tag{3.5}$$
At this modeling the acoustical resistance
$$R_A = p/Su_{osc}, \tag{3.6}$$
where $u_{osc}$ is the oscillation velocity of liquid particles, must be taken into account.

The all way from sequentially connected, via perforated holes, terms of vessel is modeled by the usual sequent resonance circuits with impedances
$$Z = \left[R_A^2 + (L_A\omega - 1/C_A\omega)^2\right]^{1/2}. \tag{3.7}$$

At the same time pressure along xilem is diminished with height as
$$p(x) \approx p_{\max} \cdot \exp\left\{-\frac{\rho cx}{R_A Z} + ik(x - ct)\right\}. \tag{3.8}$$

The most operating benefits are frequencies at which the radiation losses are minimal (resonance frequencies):
$$f_0 = \frac{1}{2\pi}\sqrt{L_A C_A} = \frac{ac}{2\pi R_0}\left[hl + 0.8ha\sqrt{\pi}\right]^{-1/2} \sim 150 \div 200 \text{ kHz}, \tag{3.9}$$
where is taken $a/R_0 = 0.5;\ l/R_0 = 0.1$.

These estimations seem the most substantiated. But they can vary for different species; for conifers, where liquid transfer between vessels take place by leakage through membrane pores, must be considered the "equivalent" pore diameters and so on. (The comparatively close of estimations deduced by several considerations including below one must be also underlined.)

Frequencies of USR generated in a crown of tree may be estimated just via (3.9) with $a$ and $l$ as radius of perforations and thickness of upper cells plates. It will leads to bigger frequencies of the order of 0.5 MHz.

But there is also more one independent possibility for estimations of phloem flux.

As was mention above there was observed a selectivity of phloem transport [9]. It must be connected with dependence of acoustic pressure on dimensions and density of moving particles [13, 14]:

$$p \sim \rho v^2 c^2 b^4 (2\pi f)^4, \qquad (3.10)$$

where $v$ and $c$ are velocities of flux and sound, $b$ is the diameter of particle. (Notice that as the acoustic field is the unique physical object that causes bigger velocity of more heavy particles, such observations gives, in principle, univocal proof of existence of the USR effects in plants.)

The effect of selectivity leads to such estimation of frequencies: energetic losses on internal friction at particles movement will be minimal at

$$f > f_{min} \approx 5 \cdot 10^{-7} b^{-2} \sim 0.5 \text{ MHz}, \qquad (3.11)$$

where for molecules of xylose, glycines, hormones is taken $b \sim 10^{-6}$ cm.

Note that this relatively power USR flux through phloem cells can be connected with the absent of nuclei in them.

Let us try to estimate the velocity of liquid flux in such USR field on the base of electroacoustical analogue.

The delay of USR in the single cell (the time duration till leaving the cell) is determined by the usual relation for harmonic oscillator [15]:

$$\tau \approx \frac{\Gamma/2}{(\omega - \omega_0)^2 + \Gamma^2/4}, \qquad (3.12)$$

where $\omega = 2\pi f$ and

$$\Gamma \equiv \frac{2}{\tau_0} = R_A C_A \omega_0^2. \qquad (3.13)$$

The minimal velocity of flux can be determined with taking into account (3.13) as

$$v_{min} = \frac{h}{\tau_0} = \frac{p}{2\rho c} \sim 1 \text{ cm/c} \qquad (3.14)$$

which is represented as reasonable even without account of viscosity.

The account of viscosity reduces this estimation. So with the Poisseuille formulae for a tree of $H' \sim 100$ m and at $\Delta p \sim 1 \div 10$ atm follows

$$v = R^2 (\rho g + \Delta p / H') / 8\eta \sim 0.1 \text{ cm/s}. \qquad (3.15)$$

This value seems consistent with other estimations.

In the well known investigation [16] Eckart determined the velocity of acoustic flow as

$$v \approx \omega p^2 / c^4 \rho^2 R^2. \qquad (3.16)$$

This expression consists at frequencies (3.9), (3.11) and a redundant pressure $p = 1 \div 10$ atm with a sap velocity of the order of 1 m/hour.

The estimation of USR power needed for such movement is the most complicate at our approach. Its part directly useful for liquids motion is of order $N_A = \rho v S$ and for a single xilem channel is of order $10^{-7}$ W, which does not seems over-big. The thermic losses can be estimated via the acoustical analog of the Bougier law

$$I(x) = I_0 \exp(-\beta x), \qquad (3.17)$$

in which for the interval $1 \div 250$ MHz is empirically usable $\beta = 10^{-17} \omega^2$ (1/cm Hz$^2$). And these values are not exorbitantly big or improbable.

## 4. Electrical phenomena accompanied liquids transport

Let's single out effects which become apparent in the conducting vessels and lead to appearance of potentials on comparatively big parts of trees stem. Their magnitudes are characterized by slow alterations with seasons, day time, atmospheric humidity, mineral fatten, e.g. [17].

These phenomena can be explainable on the base of the Debay effect [18]: in a weak electrolite ions form solvates (e.g. Na$^+ \cdot$7H$_2$O; K$^+ \cdot$6H$_2$O; Mg$^{2+} \cdot$15H$_2$O; Cl$^- \cdot$H$_2$O), which due the electrostriction have increased density and therefore acquire in the USR current higher velocity; which leads to appearing of the Debay potentials.

Inasmuch as all needed conditions are present in trees sap, the appearance of Debay potential and its alterations are their inevitable consequences. So in the dry weather a predominant lifting of alkaline ions over xilem must leads to positive charging of trees top and at high humidity, vice versa, more positive ions will go down with the phloem current: top of tree will be charged negatively (cf. [19]).

Since USR flux must go around affected parts it will wash out solvates: in such places appear the potential of affection.

Electrophysiological data gives additional possibility for estimation of USR frequency. Really, in accordance with [20] at weak alterations of potential its values are repeated at distances of order 0.5 cm along conducting ways. It evidently means the existence of wave processes with wave length of 1 cm, i.e. with the frequency of USR

$$f_{electr} \sim 200 \, \text{kHz}, \qquad (4.1)$$

completely conforms to estimations of preceding Section.

For the magnitude of Debay potential in the solution of KCl with concentration $5 \cdot 10^{-4} \div 5 \cdot 10^{-5} M$ is received $U_m = 3$ mcV on the unit of particles oscillation velocity $v = p/\rho c$ cm/c. It leads to the local pressure of USR $p = (U/U_m)\rho c$, approximately to 0.15 atm/1mcV. The observable pressure of order 10 atm can entirely stimulate the splashes of potentials of tens mV.

## 5. Some additional comments and propositions

♦5-1. USR emissions of plants are registered by some investigators (e.g. [21, 22, 23]) in the diapason of 0.1÷2 MHz, but without further analysis of spectra.

The authors tried to connect these noises with cavitation in vessels or with freezing. But even an absence of the dependence of USR intensity on wood-fibre length shows that cavitation can not be a common or unique source of noise.

♦5-2. Inasmuch as displacement of charges induces oscillation of membranes, the reciprocal effect must also exist: induced oscillation of membranes by external source on frequencies close to resonance ones would increase the speed of ions transportation into and out of cells. If the power of USR does not exceed admissible magnitudes, such influence will increase methabolism and so on. Bigger power would destroy cells. (Note that it can give key for explaining some observations of the musical and noise influence on plants.)

All these effects were observed experimentally (e.g. the review [24]). Experimentalists usually used frequencies of 400 and 800 kHz, standard for medical practice, but these frequencies are comparatively near to our estimations. For purposeful investigations the examinations with varying frequencies will be, of course, preferable.

♦5-3. The cell with opposite charges on external and internal surfaces can be considered as a spherical condensor formed by double electric layers. Such structures, as was discovered theoretically and experimentally [25, 26], emit radiowaves at alternative mechanical (acoustical) influence and vice versa, at that on the main frequencies and their harmonics (see Appendix).

All this predicts possibility of plants control by external electromagnetic radiation on resonant frequencies or their harmonics.

On the other hand it means opportunities of plant connection with each others via acoustical and/or electromagnetic fields. Such possibilities can be treated as an existence of especial "biological field" (compare e.g. [27]).

♦5-4. By existence of USR flux can be explaineted some taxises, motor activity of plants, etc.

The interesting problems are connected with organelles movement in directed fashion in mechanically perturbed cells [28]: stimulation with a glass capillary induces chloroplast migration away from the point of contact, etc. There is forward a proposal that plant membranes have stretch-activated channel activity that may be responsible for triggering ion flux changes in response to a mechanical membrane perturbation, i.e. in accordance with our hypothesis mechanical perturbation changes USR emission and corresponding field of forces.

In the ultrasound field are observed some phenomena [29], which can play definite roles in plants also. So USR in plants can promote the nitrogen fixation reactions, oppress some kinds of bacteria, etc.

♦ 5-5. Can be considered analogical phenomena in the animal cells?

Note that the more little animal cells do not have cell walls. Therefore in them would oscillate much thinner membrane, i.e. instead of $H$ in (2.4) must be taken $h$ and possibly bigger values of Young's modulus, $Y'$ instead $Y$. It leads to the estimation:

$$f_{anim} = \frac{1}{2\pi h}\sqrt{\frac{Y'}{\rho}} \sim 10^4 \div 10^5 \text{ MHz}. \tag{5.1}$$

Such frequencies (hypersound) are rapidly extinguished.

Emission of frequencies of such order was proposed in the known theory of Fröhlich, e.g. [30,31]. There are some observations in the favour of existence this radiation, e.g. [32], but the problem can not be considered as solved or closed (cf. another approach to possibilities of electromagnetic radiation [33]). Note that with plants all possible investigations seem easier and their results would be the vital for all biology.

Let's remark that there are a set of phenomena, in which can be seek, in principle, possibility of acoustical influence: hydraulic muscles of insects, motion of some protozoa and so on.

## CONCLUSIONS

Let's sum up the conducted considerations.
1. The peculiarity of external membrane of plant cells must lead to generation of chaotic ultrasound splashes by ion currents through it. Their frequencies are, most probably, in the interval 0.1 ÷ 1 MHz with intensity of tenth parts of W/cm$^2$ and induced additional pressure of the order of 1 Pa per cell.
2. This order of frequencies value follows such phenomena: 1). Ion diffusion through membrane; 2). Degassing of lifting water: 3). Resonances of xilem and phloem vessels as acoustoguides; 4). Distribution of electric potential along liquid currents. Two last estimations narrow the interval of frequencies in xilem to 150÷200 kHz.
3. Power of such generated USR seems sufficient for explanation of cyclosis, xilem and phloem transport. It explains mechanisms of external acoustical influence: acceleration of growth at small intensity and destruction at big one.
4. The existence of acoustical flows explains some electrophysiological phenomena in trees.

All it requires serious investigation of acoustical phenomena in plants by joined examinations of biologists and physicists.

## APPENDIX

Our main statement that changing of charges density on biological membranes in the processes of methabolism generates acoustical (ultrasound) waves, require more general deduction. It can be described in such form.

Changes of Hookean forces at membrane pressing on the depth $dh$ is $dp = Ydh/h$ and the complete mechanical pressure acting on the membrane

$$p_1 = Y\int \frac{dh}{h} = Y\ln\frac{h_0}{h(t)}. \tag{A.1}$$

The attractive electrical forces withstanding them are expressed by the pressure

$$p_2 = \frac{\varepsilon U^2}{2h^2} = \frac{CU^2}{2Sh}, \tag{A.2}$$

where ε is the dielectric susceptibility of membrane substance, its changing at pressing can be ignored.

The potential of membrane can be presented as

$$U(t) = U + \Delta U(t), \qquad (A.3)$$

where $U \approx 0.1$ V is the mean constant part and $\Delta U(t) < U$ is a variable part. The depth of membrane is correspondingly represented as

$$h(t) = h_0 - \Delta h_0 - x(t), \qquad (A.4)$$

here $h_0$ is the thickness of membrane without charges on it, $\Delta h_0$ is the grip at potential $U$ and $x(t)$ is the quick variation of thickness connected with $\Delta U(t)$.

Equating of (A.1) and (A.2) with taking into account the approximation $\ln(1 + \xi) \approx \xi$ at $\xi << 1$ leads to the equation

$$\frac{\Delta h + x(t)}{h(t)} = \frac{\varepsilon (U + \Delta U(t))^2}{2Y(h_0 - \Delta h - x(t))^2}. \qquad (A.5)$$

This equation can be simplified by subtracting its value at $\Delta U(t) = 0$ and $x(t) = 0$ and further division on analogical expression:

$$\frac{x(t)}{\Delta h} = 2\frac{\Delta U(t)}{U} + \left(\frac{\Delta U(t)}{U}\right)^2. \qquad (A.6)$$

This relation visually shows that each change of potential, connected with ion currents, ion absorption from exterior, molecules ionization or recombination on membranes, will generate acoustical vibrations (some close considerations are given in [34]). Note that even this form proves that at these processes along with the main frequency its second harmonics would be generated. More detail consideration would show existence of higher harmonics also.

## REFERENCES

[1]. U. Lüttge, N. Higinbotham. *Transport in plants*. NY: Springer, 1983.
[2]. H. H. Dixon, J. Joly. *Ann. Bot.* **8**, 468 (1894).
[3]. U. Zimmermann et al. *New Phytologist,* **162**, 575 (2004).
[4]. M. Th. Tyree et al. Plant Physiol., **120**, 11 (1999).
[5]. W. Konrad, A. Roth-Nebelsick. *J. Theor. Biology*, **224**, 43 (2003).
[6]. M. E. Perel'man, G. M. Rubinstein. *Biophysics*, **25**, 955 (1980) and its preprint No 482 – 80 - in Russian.
[7]. M. E. Perel'man, G. M. Rubinstein. *Bull. Acad. Sc. Georgian SSR*, **107**, 393 (1982) – in Russian.
[8]. L. D. Landau, E. M. Lifshitz. *Fluid mechanics*, Pergamon Press, New York, 1975.
[9]. P. A. Kolovsky. *Bioelectric potentials of tree plants*. Novosibirsk, Nauka, 1980 – In Russian.
[10]. K.Esau. *Anatomy of seed plants*. NY: J.Wiley. 1960.
[11]. D. T. Clarkson. *Ion transport and cell structure in plants*. L: McGraw, 1974.
[12]. H.F. Olson, *Dynamical Analogies*. NY: Van Nostrand, 1958.
[13]. L. K. Zarembo, V. A. Krasilnikov. *Introduction in Nonlinear Acoustics*. M: Nauka, 1966 - in Russian.
[14]. O. V. Rudenko and S. I. Solujan. *Theoretical foundations of nonlinear acoustics*. NY: Consult. Bureau, 1977.
[15]. E. Skudrzyk. *The foundations of acoustics*. NY: Springer, 1971.
[16]. C. Eckart. Phys, Rev., **73**, 68 (1948).
[17]. D. M. Fenson. Can. J. Bot., **41**, 831 (1963).
[18]. J. Stuehr, E.Yeager. In: *Physical acoustics. Principles and Methods* (W.P.Mason Ed.). Vol.2A. NY: Acad. Press, 1965.
[19]. A. H. De Boer, V. Volkov. *Plant, Cell and Environment*, **26,** 87 (2003).
[20]. G. La Spada et al. *Riv. Biol.*, **70**, 169 (1977).
[21]. R. T. Ritman, J. A. Milburn. *J. Exp. Bot.*, **39**, 1237 (1988)..
[22]. P. Trifilo et al. *J. Exp. Bot*, **54**, 2323 (2003)
[23]. S. B. Kikuta, H. Richter. *Plant, Cell and Environment,* **26,** 383 (2003).
[24]. D. L. Miller. *Environ. Exp. Bot.*, **23**, 1 (1983).
[25]. M. E. Perel'man, N. G. Khatiashvili. *Doklady Akademii Nauk SSSR*, **271**, 80 (1983).
[26]. N. G. Khatiashvili, M. E. Perel'man. *Phys. Earth Plan. Int.*, **57**, 169 (1989).
[27]. S. E. Shnoll et al. *Physics Uspekhi*, **41**, 1025 (1998) [*UFN*, **168**, 1129 (1998)].
[28]. J. Braam. *New Phytologist,* **165**, 373 (2005) and references therein.

[29]. I. E. El'piner, *Ultrasound: Physical, Chemical and Biological Effects*. NY: Consult. Bureau, 1964.
[30]. H. Fröhlich. *Int. J. Quantum Chem.*, **II**, 641 (1968).
[31]. H. Fröhlich. In: *Modern Bioelectrochemistry* (F. Gutman and H. Keyzer, Eds). Springer Verlag, N Y, 1986, p.14.
[32]. M. G. Akhalaya, M. S. Kakiashvili, K. A .Zakaraya, M. E. Perel'man. *Phys.Lett*. A, **101**, 367 (1984).
[33]. W. R Adey. *Physiol. Rev*., **61**, 435 (1981).
[34]. M. E. Perel'man. *Bull. Acad. Sc. Georgian SSR*, **122**, 617 (1986) – in Russian.